\documentclass{aa}  
\usepackage{epstopdf}
\usepackage{epsfig}
\usepackage{graphicx}
\usepackage{natbib}
\usepackage{gensymb}
\usepackage{txfonts}
\usepackage{amsmath}
\usepackage{dblfloatfix}
\usepackage{tabularx}
\usepackage{dblfloatfix}
\usepackage{bigints}
\newcommand*\diff{\mathop{}\!\mathrm{d}}
\newcommand{\angstrom}{\mbox{\normalfont\AA}}
\begin{document}

\title{X-ray and UV emission of the ultrashort-period, low-mass eclipsing binary system BX Tri}
\titlerunning{High-energy emission of BX Tri}

\author{V. Perdelwitz\inst{1}
  \and S. Czesla\inst{1}
  \and J. Robrade\inst{1}
  \and T. Pribulla\inst{2}
  \and J.H.M.M. Schmitt\inst{1}}

\offprints{V. Perdelwitz, \email{vperdelwitz@hs.uni-hamburg.de}}

\institute{Hamburger Sternwarte, Gojenbergsweg 112, 21029 Hamburg, Germany
\and Astronomical Institute, Slovak Academy of Sciences, 059 60 Tatransk\'a Lomnica, Slovakia}

\date{Received 22 August 2018 / Accepted 4 September 2018}
 
  \abstract
   {Close binary systems provide an excellent tool to determine stellar parameters such as radii and masses with a high degree of precision. Due to the high rotational velocities, most of these systems exhibit strong signs of magnetic activity, which has been postulated to be the underlying reason for radius inflation in many of the components.}
   {We aim to extend the sample of low-mass binary systems with well-known X-ray properties.}
   {For this, we analyze data from a singular XMM-Newton pointing of the close, low-mass eclipsing binary system BX Tri. The UV light curve is modeled with the eclipsing binary modeling tool {\sc phoebe} and data acquired with the EPIC cameras is analyzed to search for hints of orbital modulation.}
   {We find clear evidence of orbital modulation in the UV light curve and show that {\sc phoebe} is fully capable of modeling data within this wavelength range. Comparison to a theoretical flux prediction based on {\sc phoenix} models shows that the majority of UV emission is of photospheric origin. While the X-ray light curve does exhibit strong variations, the signal-to-noise ratio of the observation is insufficient for a clear detection of signs of orbital modulation. There is evidence of a Neupert-like correlation between UV and X-ray data.}
   {}


   \maketitle
%

\section{Introduction}
The increasing number of well-studied close eclipsing binary systems (EBs) with low-mass components, partly driven by large-scale exoplanet search programs (see e.g. \cite{2010Sci...327..977B} and \cite{2011AN....332..547N}), has opened up the opportunity to benchmark stellar evolutionary models such as those developed by \cite{2015A&A...577A..42B} and \cite{2012ApJ...757...42F}.
 Utilizing light curves and radial velocity data, stellar parameters such as radius and masses can be determined to a higher degree of precision than for single stars. Various authors such as \citet{2006ASPC..349...55R}, \citet{2007ApJ...660..732L} and \citet{2010ApJ...718..502M} have shown that evolutionary models can underestimate the radii of low-mass components of EBs by as much as $15\%$ and overestimate their temperatures by up to $5\%$. Although several phenomena have been proposed as the underlying reason \citep{2015ASPC..496..137F}, there seems to be a correlation between the deviations and magnetic activity \citep{2007ApJ...660..732L, 2010ApJ...718..502M, 2012ApJ...757...42F}. 
Most of these close EBs are known to be tidally locked and therefore exhibit high levels of activity caused by their rapid rotation. Based on surface magnetic flux levels published in \cite{2012LRSP....9....1R}, \cite{2013ApJ...779..183F} have computed an empirical scaling law between the X-ray luminosity $L_X$ of a star (based on ROSAT all-sky survey data \citep{1999A&A...349..389V}) and its magnetic flux $\Phi$, allowing the incorporation of magnetic fluxes derived from X-ray observations into evolutionary models.\\
 With the advent of the large X-ray facilities XMM-Newton \citep{2001A&A...365L...1J} and Chandra \citep{2000SPIE.4012....2W} it is possible to use longer pointings of EBs to derive more accurate values for $L_X$ and eliminate systematic errors such as flaring events and phase-dependent modulation. Further, in the same way that photometric light curves in the optical regime provide a precise value for the shape and extent of the photosphere, X-ray observations enable the study of coronal extent and structure previously limited to the sun \citep{2001A&A...365L.344G,2005ESASP.560..605G}.\\
\noindent
In this paper, we analyze a singular XMM-Newton observation of the close, low-mass binary system BX Tri to search for orbital modulations in the UV and X-ray regime and derive a precise estimate of the X-ray luminosity of the system. 
The paper is structured in the following way. After summarizing previous studies on the system in Section~\ref{sec:prev} we describe the observations and data products in Section~\ref{sec:analysis}. The data analysis and results are then presented in Section \ref{sec:results}, and we give a summary and conclusion of our findings in Section~\ref{sec:sc}.

\section{BX Tri: Previous knowledge}
\label{sec:prev}
BX Tri was first identified as a short-period variable star by \cite{2007A&A...467..785N} and classified as a W UMa type binary by \cite{2010MNRAS.406.2559D}, who find that it consists of a 0.51~$M_{\odot}$ primary and a 0.26~$M_{\odot}$ secondary with a separation of 1.28~$R_{\odot}$ and an orbital period of $0.19$~d, making it one of the closest known binary systems comprising of main sequence stars, below the short period limit of eclipsing binaries of 0.22~d given by \cite{2008MNRAS.388.1831R}.\\
Like most close binary systems, BX Tri exhibits strong magnetic activity in photospheric, chromospheric and coronal indicators.
Strong chromospheric activity was reported in $H\alpha$ \citep{2010MNRAS.406.2559D} as well as $H\beta$ and $H\gamma$ \citep{2014MNRAS.442.2620Z}. Both authors also find stable configurations of large spots as well as high flare rates. 
The system is identified with the RASS source 1RXS J022050.7+332049 by \cite{2007A&A...467..785N} with a count rate of $0.07\pm0.02$~ct/s and found to be a variable X-ray source by \cite{2003A&A...403..247F}. When applying the empirical conversion formula  derived by \cite{1995ApJ...450..392S} the detected ROSAT count rate corresponds to a flux of $\text{F}_\text{X}=4.4\cdot10^{-13}$~erg s$^{-1}$ cm$^{-2}$, which, when using the bolometric luminosities given in \cite{2010MNRAS.406.2559D}, corresponds to log$\left(\text{L}_\text{X}/\text{L}_{\text{bol}}\right)=-3.2$, close to the saturation limit of -3.13 \citep{2011ApJ...743...48W}. Gaia DR2 \citep{2018arXiv180409365G} states a parallax of $18.83\pm0.05$~mas (corresponding to a distance of $53.11\pm0.14$~pc) and confirms the finding of \cite{2011BlgAJ..17...39D} that BX Tri has a visual companion (** LDS 3372) at a separation of 1.2~''.
  \begin{figure}[ht!]
    \resizebox{\hsize}{!}{\includegraphics{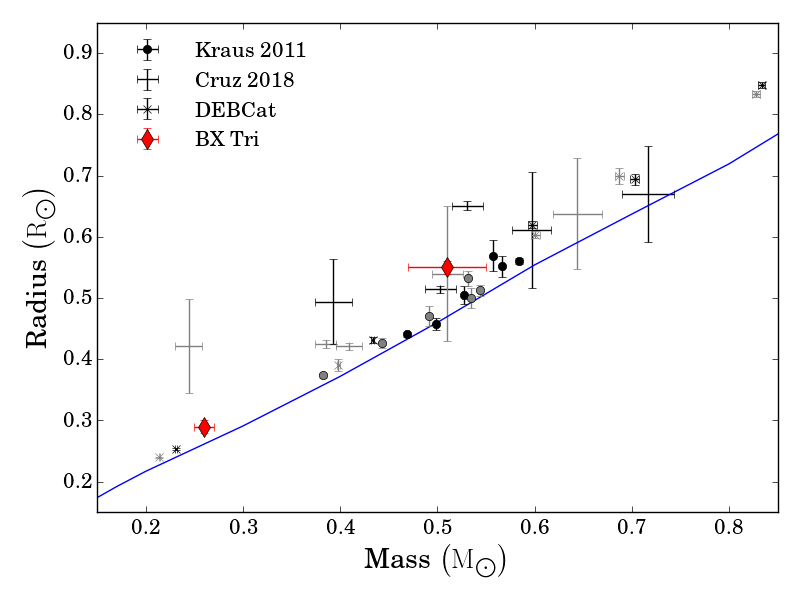}}
    \caption{Mass-radius diagram of EB components. Data are adapted from \cite{2011ApJ...728...48K}, \cite{2018MNRAS.476.5253C}, and the DEBCat catalog \citep{2015ASPC..496..164S}. Primary components are displayed as black symbols, secondaries as gray. The two components of BX Tri are displayed with the parameters determined by \cite{2010MNRAS.406.2559D}. The solid line represents a 1~Gyr isochrone model by \cite{2000ApJ...542..464C}.}
    \label{mr}
   \end{figure}
   
As is the case in many active M-Dwarfs, the radius of one of the components is enlarged substantially with respect to model predictions. Figure~\ref{mr} shows the mass-radius relationship of close binary systems based on data from \cite{2011ApJ...728...48K}, \cite{2015ASPC..496..164S} and \cite{2018MNRAS.476.5253C}, as well as both components of BX Tri,  assuming the values derived by \cite{2010MNRAS.406.2559D}. A comparison to a 1~Gyr isochrone model \citep{2000ApJ...542..464C} shows that most stars within the sample are significantly enlarged with respect to model predictions.\\
\cite{2013A&A...549A..86L} found period variations of $\text{dP}/\text{dt}=(-0.030\pm0.007)$~s/yr\footnote{The updated values are from priv. comm.} with a 4~$\sigma$ significance level based on SuperWASP data \citep{2006PASP..118.1407P}.
The system parameters derived by \cite{2010MNRAS.406.2559D} and \cite{2014MNRAS.442.2620Z} are listed in Table~\ref{tab1}.\\
\begin{table}[htb!]
\caption{System parameters of BX Tri. The second column lists the values derived by \cite{2010MNRAS.406.2559D}, the third those by \cite{2014MNRAS.442.2620Z} for a semi-detached configuration.}         
\label{table:1} 
\centering    
\begin{tabular}{c c c}       
\hline\hline                      
Parameter &  DK2010 & Z2014\\   
\hline                 \\ 
    $P$ [d] 		&  0.192637 		& 0.19263595\\  
    $T_1$ [K] 		& 3735 $\pm$ 10      	& 3735 $\pm$ 10\\
    $T_2$ [K] 		&  3106 $\pm$ 10        & 3359 $\pm$ 28\\
    $M_1$ [$M_\odot$] 	&0.51 $\pm$ 0.02     	& 0.578 $\pm$ 0.04\\
    $M_2$ [$M_\odot$] 	& 0.26 $\pm$ 0.02	& 0.280 $\pm$ 0.02\\
    $R_1$ [$R_\odot$] 	& 0.55 $\pm$ 0.01     	& 0.59 $\pm$ 0.01\\
    $R_2$ [$R_\odot$] 	& 0.29 $\pm$ 0.01     	& 0.27 $\pm$ 0.01\\
    $L_1$ [$L_\odot$] 	& 0.053 $\pm$ 0.002     & - \\
    $L_2$ [$L_\odot$] 	& 0.0070 $\pm$0.0006    & - \\
    $i[^\circ]$ 	& $72.5$ 		& 66.89 $\pm$ 0.45\\
    $a$ [$R_\odot$] 	& 1.28 $\pm$ 0.04     & 1.33 $\pm$ 0.03\\
    $d$ [pc]		& 59 $\pm$ 2 		& -\\ \\ 
\hline                                          
\end{tabular}
\label{tab1}
\end{table}

\section{Observations}
\label{sec:analysis}
\begin{figure}[!htb]
\includegraphics[width=0.48\textwidth]{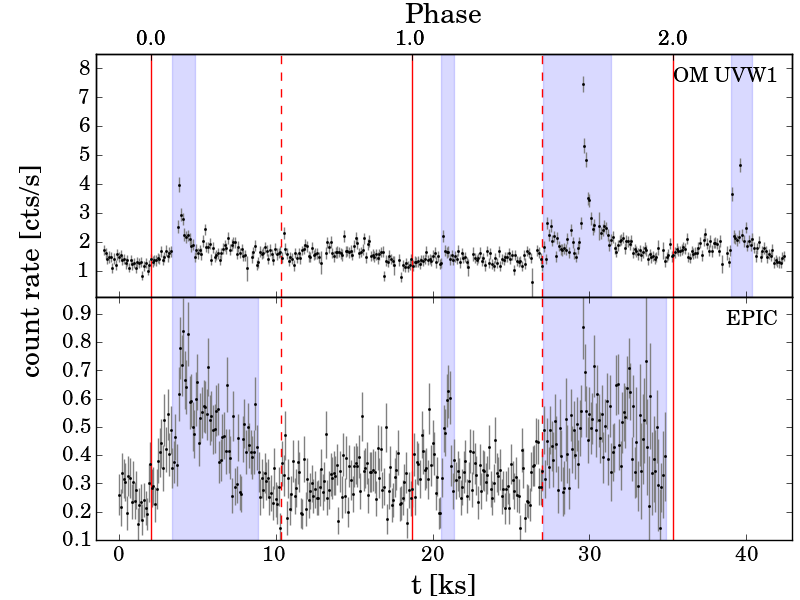}
\caption{XMM-Newton light curves of BX Tri. Upper panel: OM UVW1 data. Lower panel: Co-added EPIC light curves for the energy range 0.2-5~keV. The position of the primary and secondary eclipses based on the ephemerides given by \cite{2014MNRAS.442.2620Z} are marked by solid and dashed vertical red lines. Flaring regions excluded from the light curve analysis described in Sections~\ref{sec:UV} and \ref{sec:X} are marked with shaded areas.}
\label{fig:lc}
\end{figure}

Observations of BX Tri were carried out with XMM-Newton on 2013-07-19 (Obs. ID 0720180101) for a duration of $43$~ks with the Optical Monitor (OM) and $35$~ks with the EPIC detectors, covering two full periods of the system. All data were reduced using the XMM-Newton {\sc science analysis system (sas)} version 14.0.0. and barycentric correction was carried out.\\ 
All EPIC cameras were operated in full frame mode using the medium filter. In order to detect eclipses or other signs of periodicity, light curves were extracted for an energy range of $0.2-5$~keV for all EPIC detectors and binned to $100$~s time intervals. Throughout the exposure the background level was substantial, but particularly towards the end of the pointing, the X-ray data quality deteriorated significantly and all EPIC detectors switched off 35~ks into the observation.\\ 
Since the Reflection Grating Spectrometer (RGS) spectra exhibit a low signal-to-noise ratio, spectroscopy was performed with the EPIC detectors only.\\
The XMM-Newton OM \citep{2001A&A...365L..36M} observed in fast mode setting using the UVW1 filter (effective wavelength/width $291/83$~nm) in order to acquire data in the ultraviolet. Data reduction was performed with the SAS OM pipeline {\sc omfchain} and light curves were binned to 60~s intervals.\\
Figure~\ref{fig:lc} shows the light curves of the entire observation in the UV (upper panel) and X-ray range (lower panel). Both wavelength regimes exhibit significant flaring events, with the count rates rising by a factor of 7 in the UV and 4 in the X-ray. In Figure~\ref{fig:lc} these flaring regions, determined by visual inspection, are marked with shaded areas, along with solid and dashed lines representing the locations of primary and secondary eclipses.
 A first look at the OM light curve gives the impression that primary eclipses are present at phases $0.0$ and $1.0$, while the third at phase $2.0$ is obscured by a fading flare. The secondary eclipses, however, appear to be indistinguishable from noise.\\
A similar inspection of the X-ray light curve yields no obvious signs of orbital modulation.

\section{Data analysis and Results}
\label{sec:results}
In this Section we discuss the analysis of the UV and X-ray light curves as well as EPIC spectra. For this purpose, both the UV and X-ray light curves were phase-folded according to the ephemerides derived by \cite{2010MNRAS.406.2559D}, and the orbital modulations in both wavelength regimes were utilized to draw conclusions on the origin of the UV emission as well as the distribution of the coronal plasma.
We then conclude the section with a discussion of the flaring events.
\subsection{UV range}
\label{sec:UV}
\begin{figure}[!hbt]
\includegraphics[width=0.48\textwidth]{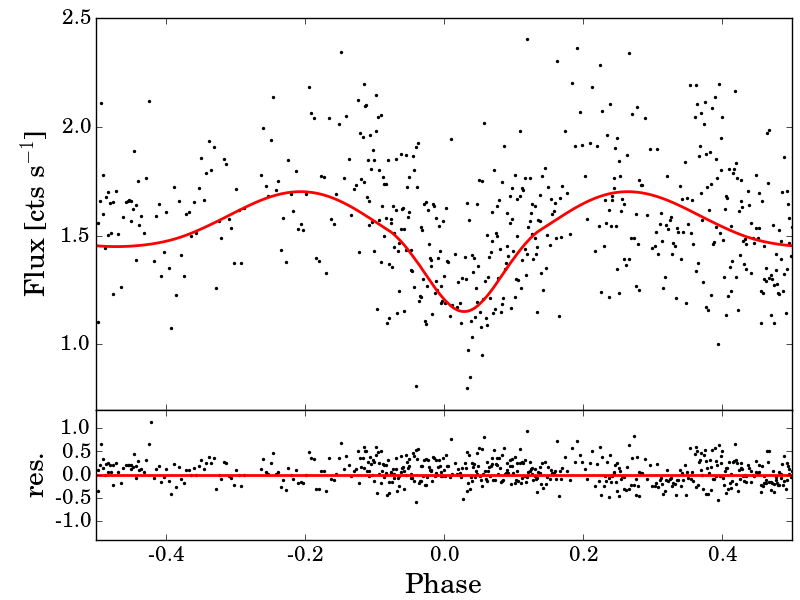}
\caption{OM light curve folded with the period given by \citep{2014MNRAS.442.2620Z}. The solid red line represents the {\sc phoebe} model and the residuals between the flux and the unshifted model are displayed in the bottom panel. Regions with flaring events were omitted.}
\label{fig:fit}
\end{figure}

As a first approach to analyzing the UV light curve we compare it to model predictions based on the {\sc phoebe} code \citep{2005ApJ...628..426P}. 
Optical light curves and RV data of BX Tri from \cite{2010MNRAS.406.2559D} were fitted following the procedure described in their publication, and subsequently removing the two spots from the model. 
The OM UVW1 light curve was then added as an additional data set, omitting the flaring regions marked as shaded areas in Figure~\ref{fig:lc}. For this purpose, we added the XMM-Newton OM UVW1 filter~\footnote{\url{ftp://xmm.esac.esa.int/pub/ccf/constituents/extras/responses/OM/}} to the set of {\sc phoebe} filter curves. 
Since the van Hamme limb darkening coefficients \citep{1993AJ....106.2096V} utilized by {\sc phoebe} cover the spectral range in question, we could then compare the model predictions to the observed UV light curve. The resulting light curve, along with {\sc phoebe} model predictions, is displayed in Figure~\ref{fig:fit}.
While the residuals show slight systematics around the primary eclipse, the artificial light curve yields a good approximation of the data, indicating that the modeling of eclipses in the UV with {\sc phoebe} works well. The flux in the UVW1 filter at quarter phase based on the model light curve was determined to be $(6.72\pm0.04)\times10^{-13}$~erg/s/cm$^2$, the flux during primary eclipse to be $(4.54\pm0.01)\times10^{-13}$~erg/s/cm$^2$. Here, we used the conversion factor of $4.76\times10^{-16}$~erg/cm$^2$/$\angstrom$/cnt as stated on the {\sc SAS} website\footnote{\url{https://www.cosmos.esa.int/web/xmm-newton/sas-watchout-uvflux}}. 
A geometrical model assuming two spheres yields a primary eclipse depth of around $10\%$, which is below the observed variation, indicating that Roche geometry and heating effects play an important role in the modeling of the system.\\ 
We compare the out-of-eclipse flux to a theoretical flux prediction based on a {\sc phoenix} \citep{1999ApJ...512..377H} composite model with temperatures $T_1/T_2=3700/3100$~K, surface gravities  log~$g=4.5/5.0$ and solar metallicities. The model spectra $S(\lambda)$ of primary and secondary component are multiplied with the respective stellar surfaces and folded with the OM effective area.
The theoretical flux prediction at quarter phase can then expressed as
\begin{equation}
F_q=\frac{\bigintsss_{\;0}^{\infty}\!\left(S_p(\lambda)\,A_p+S_s(\lambda)\;A_s\right)\times\Phi_{eff}(\lambda)\diff\lambda}{4\pi d^2\bigintsss_{\;0}^{\infty}\Phi_{eff}(\lambda)\diff\lambda}\times\sigma_{eff}
\end{equation}
where $A_{p/s}=4\pi R_{p/s}^2$ are the surfaces of the primary and secondary component, $\Phi_{eff}(\lambda)$ is the effective area of the UVW1 filter \citep{2009Ap&SS.320..177T}, $d=53.11$~pc is the distance of BX Tri and $\sigma_{eff}=830\angstrom$ is the effective width of the filter.\\
This approximation yields an out-of-eclipse flux of $F_q=5.20\times10^{-13}$~erg/s/cm$^2$, which corresponds to $77\%$ of the measured flux. Since this approach neglects Roche geometry and surface heating effects, both of which should increase the predicted flux, we conclude that at least three quarters of the UV flux is of photospheric origin, since the {\sc phoenix} models do not include chromospheric emission.  
Since both the {\sc phoebe} model and the {\sc phoenix} spectra give a good approximation of the observed UV flux, we assume that both components contribute to the UV emission, despite the absence of clear secondary eclipses in the light curve.

\begin{figure}[!htb]
\includegraphics[width=0.48\textwidth]{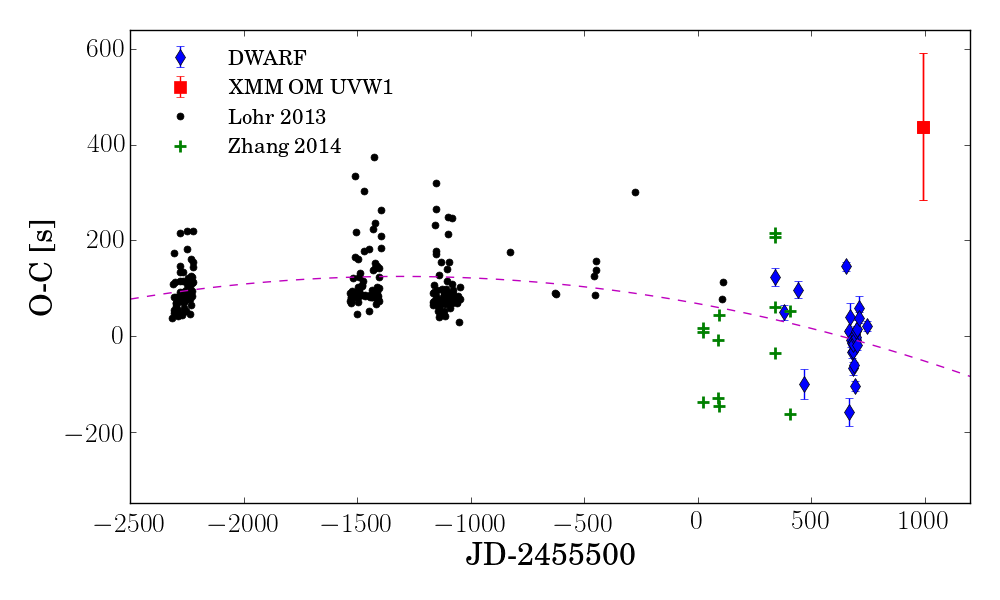}
\caption{Observed-Calculated eclipse minima of BX Tri. Archival data from \cite{2014MNRAS.442.2620Z} are marked with green crosses, those from and \cite{2013A&A...549A..86L} and Lohr (priv. comm.) with black dots. The DWARF project data \citep{2012AN....333..754P} are displayed as blue diamonds. The XMM-Newton OM minimum is highlighted in red. The best-fit quadratic function based on the visual data is displayed as a broken line and yields a period change of $(-0.018\pm0.003)$~s/yr.}
\label{fig:OC}
\end{figure}
Using a least-squares analysis, we determined the phase of the primary eclipse in the folded light curve.
Assuming the ephemerides given in \cite{2014MNRAS.442.2620Z}, we find a substantial phase shift in the OM data of $0.029\pm0.009$~phases corresponding to $(483\pm153)$~s. Here, the error was determined using a Monte-Carlo approach, displacing the flux values within the respective error range and determining the best-fit phase shift for each draw. The standard deviation of the phase shifts was adopted as the error value. In order to compare this phase shift to O-C data acquired in the optical regime, we combine data derived by  \cite{2013A&A...549A..86L}, \cite{2014MNRAS.442.2620Z} and \cite{2012AN....333..754P}. The resulting O-C diagram is displayed in Figure~\ref{fig:OC}, along with a quadratic fit of all available optical data. 
We find a significant period change of $(-0.024\pm0.005)$~s/yr in the optical O-C data, which is consistent with an updated value of $(-0.030\pm0.007)$~s/yr derived by \cite{2013A&A...549A..86L}\footnote{We obtained the updated value, based on the analysis of additional data, from the authors by private communication.}. The O-C value of the UV minimum does not fit the trend of the optical data within the margins of error.
This indicates that the emitting material is not distributed in a spherically symmetric way across the surface of each stellar component, which is in agreement with the fact that \cite{2010MNRAS.406.2559D} find large spots in the photospheres of the binary.

\subsection{X-ray range}
\label{sec:X}
\begin{figure}[!htb]
\centering
\includegraphics[width=0.48\textwidth]{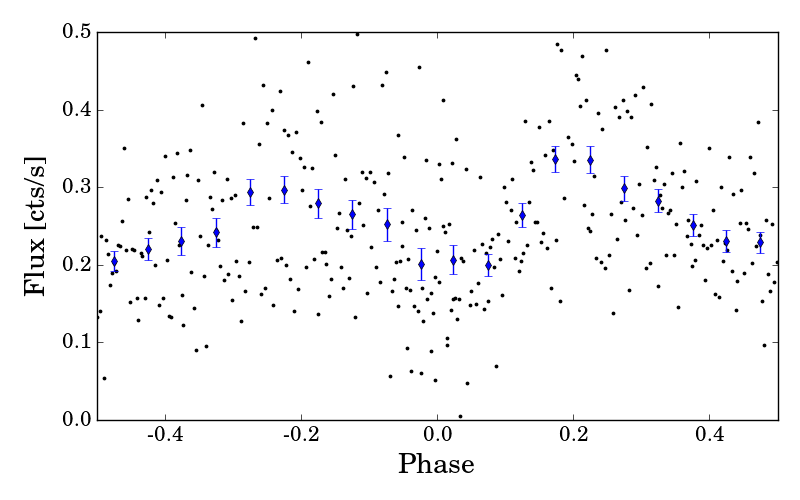}
\caption{Complete EPIC PN light curve of BX Tri folded with the period derived by \cite{2014MNRAS.442.2620Z}. The black dots represent the full light curve with a binning of 100~s (error bars were omitted for clarity), while the blue diamonds display the phase-binned light curve.}
\label{fig:xray_lc_1}
\end{figure}

\begin{figure}[!htb]
\centering
\includegraphics[width=0.48\textwidth]{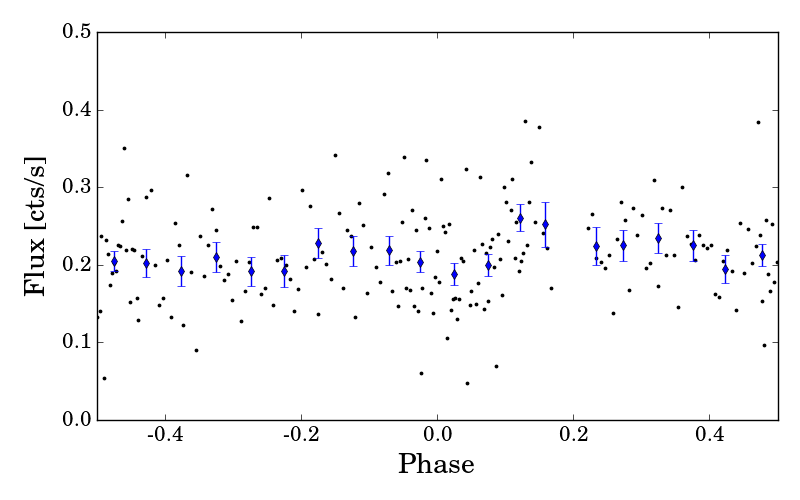}
\caption{EPIC PN light curve of BX Tri folded with the period derived by \cite{2014MNRAS.442.2620Z} after flare removal. The black dots represent the filtered light curve with a binning of 100~s (error bars were omitted for clarity), while the blue diamonds display the phase-binned light curve.}
\label{fig:xray_lc_2}
\end{figure}

As a first attempt to identify orbital modulation, we phase-folded the entire EPIC PN light curve. While the result, displayed in Figure~\ref{fig:xray_lc_1}, shows eclipse-like modulations at the location of the primary and secondary eclipse, a closer look at the lower panel of Figure~\ref{fig:lc} reveals that this impression is caused by two large flaring events. An increase in flare rates at quarter-phase has been observed in the optical regime in other binary systems \citep{2016ApJS..224...37G}.\\
The same light curves were then phase-folded after removal of the large flaring events by visual inspections (shaded regions in Figure~\ref{fig:lc}). The resulting light curve is shown in Figure~\ref{fig:xray_lc_2} and does not exhibit any clear signs of eclipses.\\
If we assume both stars to be magnetically saturated ($L_X/L_{bol}=10^{-3.13}$), the distribution of the coronal emission then follows the ratio of the bolometric luminosities, or $L_X^1/L_X^2=L_{bol}^1/L_{bol}^2$. Even under optimal conditions for the detection of eclipses, that is with no coronal extent above the photosphere, the depth of primary and secondary eclipse would be of the order of $0.015$~cts/s and $0.007$~cts/s, respectively, which is well below the noise level of the light curve.\\ 
For an estimate of the coronal scale height, a higher signal-to-noise ratio would be required.\\ 
Without the presence of eclipses, we cannot determine the contribution of each stellar corona to the overall X-ray luminosity. Based on the fact that both components are extremely fast rotators we can make the reasonable assumption that the secondary component also exhibits coronal activity and contributes to the X-ray emission.

\begin{figure}[!ht]
\centering
\includegraphics[width=0.48\textwidth]{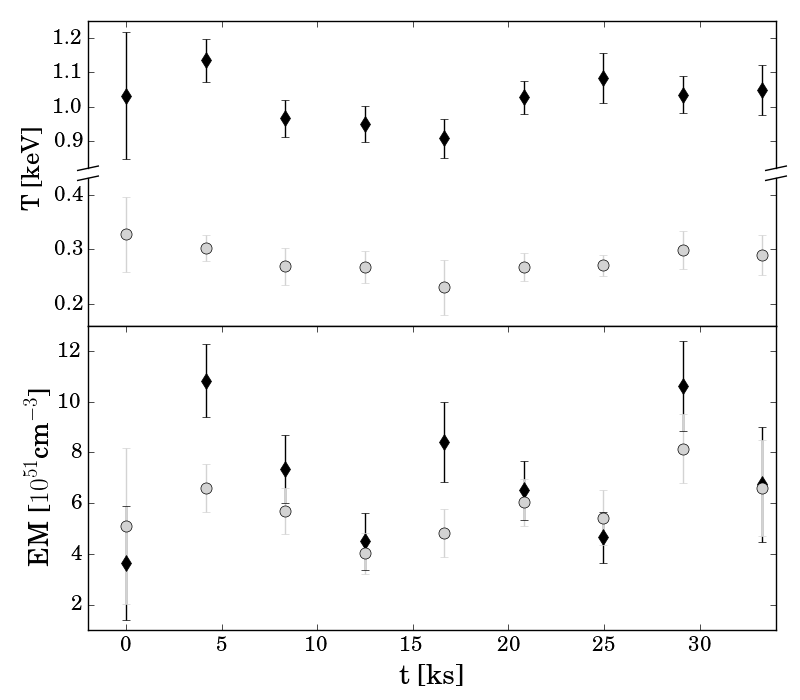}
\caption{Plasma temperatures (upper panel) and emission measures (lower panel) derived from time-resolved spectroscopy. The black diamonds represent the higher, the gray circles the lower temperatures of the two-temperature apec model fit.}
\label{fig:spec}
\end{figure}

\subsubsection{Time-resolved spectroscopy}
We divided the EPIC data into 9 time intervals centered on times corresponding to quarter phases and extracted the individual spectra. Spectral analysis was performed with {\sc xspec} version 12.8.1 \citep{1996ASPC..101...17A} by simultaneously fitting MOS1, MOS2 and PN spectra with an apec model with two plasma temperatures.
\begin{table*}[!b]
\caption{Results of the time resolved spectroscopy. The data were centered on the quarterly phases stated in column one, resulting in the time-spans given in column 2. The temperatures and emission measures of the $2-T$ apec model fit are displayed in columns 3-6. The last column shows the X-ray luminosities in the range 0.2-10~keV.}              
\label{table:2}      
\centering                                      
\begin{tabular}{c c c c c c c}          
\hline\hline                        
Phase&time& T$_1$&EM$_1$ & T$_2$&EM$_2$&$\text{L}_\text{X}$\\
& [ks] & [keV]&[$10^{51}$cm$^{-3}]$& [keV]&[$10^{51}$cm$^{-3}]$&[$10^{28}$~erg/s]\\
\hline                         
0&0.0-4.2&$1.03\pm0.19$&$3.65\pm2.24$&$0.33\pm0.07$&$5.09\pm3.08$&$9.51\pm0.21$\\
0.25&4.2-8.3&$1.14\pm0.06$&$10.82\pm1.44$&$0.30\pm0.02$&$6.60\pm0.93$&$17.31\pm0.08$\\
0.5&8.3-12.5&$0.97\pm0.05$&$7.34\pm1.33$&$0.27\pm0.03$&$5.70\pm0.91$&$10.92\pm0.10$\\
0.75&12.5-16.6&$0.95\pm0.05$&$4.49\pm1.11$&$0.27\pm0.03$&$4.02\pm0.81$&$9.29\pm0.08$\\
1&16.6-20.8&$0.91\pm0.06$&$8.41\pm1.57$&$0.23\pm0.05$&$4.83\pm0.95$&$10.19\pm0.09$\\
1.25&20.8-25.0&$1.03\pm0.05$&$6.50\pm1.15$&$0.27\pm0.03$&$6.03\pm0.93$&$11.70\pm0.10$\\
1.5&25.0-29.1&$1.08\pm0.07$&$4.66\pm1.01$&$0.27\pm0.02$&$5.43\pm1.10$&$9.96\pm0.09$\\
1.75&29.1-33.3&$1.03\pm0.05$&$10.62\pm1.77$&$0.30\pm0.03$&$8.14\pm1.35$&$15.79\pm0.07$\\
2&33.3-37.4&$1.05\pm0.07$&$6.74\pm2.25$&$0.29\pm0.04$&$6.60\pm1.89$&$15.08\pm0.20$\\
\hline                                             
\end{tabular}
\end{table*}
Table~\ref{table:2} contains the center phase of the time intervals, the time span relative to the start of the observation, as well as temperatures and emission measures of both plasma components. The reduced $\chi^2$ of all spectral fits was below 1.2, indicating that the two-temperature plasma model describes the coronal emission well.\\
Figure~\ref{fig:spec} shows the plasma temperatures and emission measures derived from time-resolved spectroscopy, which exhibit no signs of periodicity, but rather follow the trend of the X-ray light curve, a fact which we attribute to the presence of large flares during the observation.\\ 
We determined the quiescent X-ray luminosity by extracting the spectrum for the region between the first secondary and second primary eclipse (corresponding to phase 0.75 in Table~\ref{table:2}), which is unaffected by larger flares, yielding L$_\text{X} = (9.29\pm0.07)\cdot10^{28}$~erg/s in the energy range $0.2-5$~keV, or $\text{log}(L_X/L_{bol})$=-3.4.

\subsection{Flaring events}
\cite{2014MNRAS.442.2620Z} state a flare rate of BX Tri of $\nu_f=0.11/h$ in the optical regime, which among close, low-mass EBs is only surpassed by YY Gem with $\nu_f=0.22/h$ \citep{1990A&A...227..130D}. During the XMM-Newton observation of BX Tri described in this paper, three large and several small flaring events occurred.
All of these appear first in the UV light curve and are then echoed in the X-ray regime, where the decay time is much larger, which is characteristic of the X-ray emitting material filling the coronal loops.\\ 
In order to obtain an estimate of the emitting area and total energy of the two flares ($t_1=3.4$~ks and $t_2=29$~ks after the start of the EPIC observations) we followed the approach of \cite{2013ApJS..209....5S} and estimated the area as
\begin{equation}
A_{flare}=C'_{flare}\,\pi\,\frac{\bigintsss_{\;0}^{\infty} R_{\lambda}\,\left(R_1^2\,B_{\lambda}(T_1)+R_2^2\,B_{\lambda}(T_2)\right)\,d\lambda}{\bigintsss_{\;0}^{\infty} R_{\lambda}\,B_{\lambda}(T_{flare})\,d\lambda}
\end{equation}
where $C'_{flare}$ is the ratio of the flare peak and quiescent count rates, $R_i$ are the stellar radii, $R_{\lambda}$ is the XMM-Newton OM response curve and $B_{\lambda}(T_i)$ are the Planck curves at the effective and flare temperature. Here, we used the value of $T_{flare}=9000$~K stated by \cite{2011A&A...530A..84K} as the flare temperature. With count rate ratios of $C'_1=2.7$ and $C'_2=4$, we estimated the flare area to be $A_{flare}\approx4.8\times10^{18}$~cm$^2$, or $0.08\%$ of the combined stellar surface, for the first flare and $A_{flare}\approx1.1\times10^{19}$~cm$^2$, or $0.17\%$ of the combined stellar surface, for the second.\\
We could then derive the bolometric flare luminosities via
\begin{equation}
 L^{flare}=\sigma\,T^4_{flare}\,A_{flare},
\end{equation}
where $\sigma$ is the Stefan-Boltzmann constant, and arrived at peak luminosities of $L^{flare}_1=1.8\times10^{30}$~erg/s and $L^{flare}_2=4.1\times10^{30}$~erg/s.\\
\begin{figure}[!ht]
\centering
\includegraphics[width=0.48\textwidth]{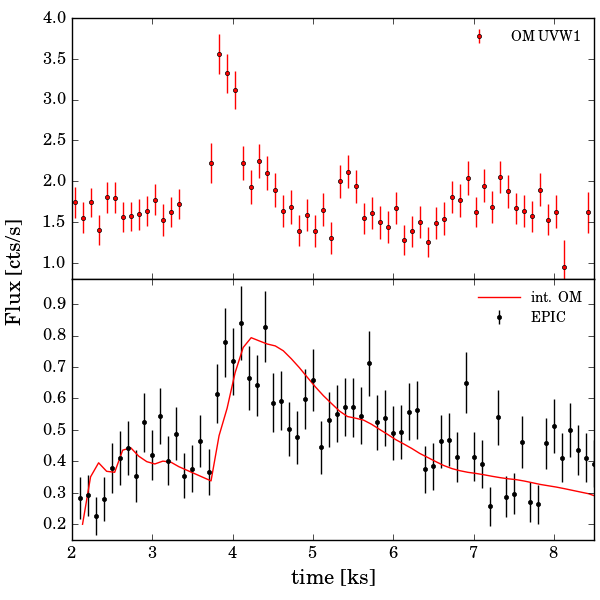}
\caption{Flaring event at $t=3.4$~ks in the UVW1 filter (upper panel) and the X-ray regime (lower panel). The solid line in the lower panel represents the modeling of a Neupert-like effect according to Equation~\ref{eq:1}.}
\label{fig:neupert}
\end{figure}
Finally, we investigated the existence of a Neupert-like effect. The Neupert effect \citep{1968ApJ...153L..59N} originally described a correlation between microwave and soft X-ray emission. It is not only of interest with regard to flare energetics, but has also been proposed by \cite{1988ApJ...330..474P} as a contributor to coronal heating.\\ 
Our effort was limited by the high background level and the read-out gaps in the OM data, and we limited it to the flaring event occuring 3.4~ks after the beginning of the observation. We followed the formalism of \cite{1996ApJ...471.1002G} and compared the generalized Neupert effect
\begin{equation}\label{eq:1}
 L(t)=\frac{\alpha}{\tau(t)}\int_{t_0}^{t}F_{UV}e^{-(t-u)/\bar{\tau}(t,u)}du,
\end{equation}
where L denotes the X-ray luminosity, $\alpha$ is a proportionality constant, $\tau$ is the thermal decay parameter, and $F_{UV}$ is the flux in the UVW1 filter.
Figure~\ref{fig:neupert} shows the UVW1 light curve (upper panel), the X-ray light curve (black dots in lower panel) as well as the X-ray count rate derived via Equation~\ref{eq:1} (red solid line in lower panel). While the derived X-ray light curve follows the general trend of the observed data, the quality of the approximation is diminished by the occurrence of smaller flares after the large peak, as well as by a read-out gap in the OM data at the beginning of the flare.
Due to this fact and the low count rates in the X-ray regime it is not possible to study delay times between UV and X-ray flares in a manner similar to \cite{2005A&A...431..679M}.\\

\section{Summary and conclusions}
\label{sec:sc}
We have performed a detailed analysis of data from a XMM-Newton observation of the close, low-mass eclipsing binary system BX Tri. Based on the OM fast mode data acquired with the UVW1 filter, we were able to model the UV emission with {\sc phoebe} and conclude that at least three quarters of the UV flux originates in the stellar photosphere. A comparison with optical O-C data indicates an inhomogeneous distribution of emitting material likely caused by spot modulation.\\
In the X-ray regime we found that the signal-to-noise ratio was insufficient to detect orbital modulation and derive a unique solution of the coronal distribution and scale heights. In comparison to the close M-dwarf binary YY Gem, which exhibits clear eclipses in the X-ray regime \citep{2001A&A...365L.344G,2002ASPC..277..215S}, the X-ray luminosity and orbital inclination of BX Tri are less favorable for such an analysis.\\
The quiescent, out-of-eclipse X-ray luminosity of the system was determined to be L$_\text{X} = (9.29\pm0.07)\cdot10^{28}$~erg/s, or $\text{log}(L_X/L_{bol})=-3.4$, placing BX Tri near the magnetic saturation limit of $\text{log}(L_X/L_{bol})=-3.13$.
We find that the large flaring events during the observation dominate both the X-ray light curve and the parameters of the coronal plasma derived via time-resolved spectroscopy. We determined that there is evidence of a Neupert-like relation between the UV and X-ray light curve and estimated the X-ray luminosities of two larger flaring events.\\
While secondary eclipses are absent in the UV and X-ray light curves, we stipulate that both stellar components contribute to the observed flux in the two wavelength regimes.

\begin{acknowledgements}
Based on observations obtained with XMM-Newton, an ESA science mission with instruments and contributions directly funded by ESA Member States and NASA. This work has made use of data from the European Space Agency (ESA) mission
{\it Gaia} (\url{https://www.cosmos.esa.int/gaia}), processed by the {\it Gaia}
Data Processing and Analysis Consortium (DPAC,
\url{https://www.cosmos.esa.int/web/gaia/dpac/consortium}). Funding for the DPAC
has been provided by national institutions, in particular the institutions
participating in the {\it Gaia} Multilateral Agreement.
VP acknowledges funding through the DFG Research Training Group 1351.
JR acknowledges funding through DLR grant 50QR1605.
SC acknowledges support from DFG project SCH 1382/2-1
This work was supported by project VEGA 2/0031/18.
We thank M.E. Lohr for supplying us with updated data from the SuperWASP analysis.
\end{acknowledgements}

\bibliographystyle{aa} 
   \bibliography{biblio.bib} 
%

\end{document}